\documentclass{ws-procs9x6-fix}
\usepackage{graphicx}

\newcommand{\bc}{\begin{center}}
\newcommand{\ec}{\end{center}}
\newcommand{\be}{\begin{equation}}
\newcommand{\ee}{\end{equation}}
\newcommand{\bea}{\begin{eqnarray}}
\newcommand{\eea}{\end{eqnarray}}

\begin{document}

\title{From Equilibrium to Transport Properties of\\ 
Strongly Correlated Fermi Liquids}

\author{Thomas Sch\"afer}

\address{Department of Physics, North Carolina State University, \\
Raleigh, NC 27695}

\begin{abstract}
We summarize recent results regarding the equilibrium 
and non-equilibrium behavior of cold dilute atomic gases 
in the limit in which the two body scattering length $a$
goes to infinity. In this limit the system is described
by a Galilean invariant (non-relativistic) conformal
field theory. We discuss the low energy effective lagrangian 
appropriate to the limit $a\to\infty$, and compute low
energy coefficients using an $\epsilon$-expansion. We 
also show how to combine the effective lagrangian with 
kinetic theory in order to compute the shear viscosity,
and compare the kinetic theory predictions to experimental 
results extracted from the damping of collective modes in 
trapped Fermi gases.

\end{abstract}


\bodymatter

\section{Introduction}
\label{sec_intro}

  Over the last ten years there has been remarkable progress
in the study of ``designer fluids'', dilute, non-relativistic Bose
and Fermi gases in which the scattering length between the
Bosons or Fermions can be continuously adjusted. In the following
we are particularly interested in Fermi gases, since these 
systems are stable for both positive and negative values 
of the scattering length, including the strongly correlated
limit in which the scattering length is taken to infinity. 

 The scattering length is controlled through a Feshbach resonance. 
Alkali atoms such as $^6$Li and $^{40}$K have a single valence electron. 
When a dilute gas of atoms is cooled to very low temperatures, we can 
view the atoms as pointlike particles interacting via interatomic 
potentials which depend on the hyperfine quantum numbers. A Feshbach resonance 
arises if a molecular bound state in a ``closed'' hyperfine channel crosses 
near the threshold of an energetically lower ``open'' channel. Because 
the magnetic moments of the open and closed states are in general 
different, Feshbach resonances can be tuned using an applied magnetic 
field.  At resonance the two-body scattering length in the open channel 
diverges, and the cross section $\sigma$ is limited only by unitarity, 
$\sigma(k) = 4\pi/k^2$ for low momenta $k$. In the unitarity limit, 
details about the microscopic interaction are irrelevant, and the 
system displays universal properties.

 Near a Feshbach resonance the scattering length behaves as
\be 
a = a_0 \left( 1 + \frac{\Delta B}{B-B_0}\right) \,
\ee
where $a_0$ is the non-resonant value of the scattering length 
(typically on the order of the effective range of the interatomic
potential), $B$ is the magnetic field, $B_0$ the position of the
resonance, and $\Delta B$ the width. A small negative scattering 
length corresponds to a weak attractive interaction between
the atoms. This case is known as the BCS (Bardeen-Cooper-Schrieffer)
limit. On the other side of the resonance the scattering length is 
positive. In the BEC (Bose-Einstein condensation) limit the interaction 
is strongly attractive and the fermions form deeply bound molecules.
For this reason the unitarity limit $a\to\infty$ is also known 
at the BCS/BEC crossover. 

 The unitarity limit is of interest to QCD practitioners for a
for a number of reasons:

\begin{itemize}

\item{The unitarity limit provides an approximate description of 
dilute neutron matter. The neutron-neutron scattering length 
is $a_{nn}=-18$ fm, and the effective range is $r_{nn}=2.8$ fm. 
This means that there is a range of densities, relevant to the 
outer layers of neutron stars, for which the interparticle
spacing is large compared to the effective range, but small
compared to the scattering length.}

\item{The Fermi gas at unitarity is a high $T_c$ superconductor. 
There is an attractive interaction in the spin singlet channel
which leads to s-wave superconductivity below some critical 
temperature $T_c$. In the unitarity limit the only energy 
scale in the problem is the Fermi energy $E_F$, and we must 
have $k_BT_c=\alpha E_F$ with some numerical constant $\alpha$. 
Quantum Monte Carlo calculations (and experimental results)
indicate that $\alpha\simeq 0.15$ \cite{Burovski:2006,Bulgac:2008}, 
much larger than in ordinary (or even high $T_c$) electronic 
superconductors, but comparable to what might be achieved in 
color superconducting quark matter \cite{Alford:2007xm}.}

\item{The limit $a\to\infty$ corresponds to a non-relativistic
conformal field theory \cite{Mehen:1999nd}. In the unitarity limit 
there is no scale in the problem (other than the thermodynamic 
variables temperature and density). Indeed, one can show that the 
theory is not only scale invariant, but invariant under the full 
conformal group. This raises the question whether there are any 
physical consequences of conformal symmetry that go beyond results
that follow from scale invariance. It also raises the possibility
that a holographic description, similar to the $AdS/CFT$ correspondence,
can be obtained \cite{Son:2008ye,Balasubramanian:2008dm}.}

\item{Non-relativistic fermions at unitarity behave as a very
good fluid and show interesting transport properties, including 
a very small shear viscosity. Kinetic theory suggests that the 
shear viscosity is inversely proportional to the scattering 
cross section, and reaches a minimum at unitarity. This expectation
is confirmed by experiments that demonstrate large elliptic flow 
and a very small damping rate for collective oscillations
\cite{oHara:2002,Kinast:2005}. }
\end{itemize}

\section{Equilibrium Properties}
\label{sec_equ}

 We begin by analyzing equilibrium properties of the dilute 
Fermi gas at unitarity. If the temperature is large, $k_BT>E_F$, 
then the scattering cross section is regularized by the thermal
wave length, and the effective interaction is weak. Here the Fermi
energy is defined by $E_F=(3\pi^2 n)^{2/3}/(2m)$, where $n$ is the 
density, and $m$ is the mass of the atoms. In the high temperature 
regime the equation of state is well described by the Virial expansion, 
and the system has single particle excitations with the quantum numbers 
of the fundamental fermions. In the regime $k_BT\sim E_F$ the interactions 
are strong. As noted above, superfluidity occurs at $k_BT_c\simeq 
0.15 E_F$. Below the critical temperature the excitations are Goldstone
bosons. In following section we will discuss the effective 
theory of the Goldstone bosons, and relate the parameters in the
effective lagrangian to static properties of the system. 

\subsection{Low Energy Effective Theory and Density Functional}
\label{sec_let}

   The Goldstone boson field can be defined as the phase of the 
difermion condensate $\langle\psi\psi\rangle = e^{2i\varphi}
|\langle \psi\psi\rangle|$. The effective Lagrangian at next-to-leading
order (NLO) in derivatives of $\varphi$ and the external potential 
is \cite{Son:2005rv}
\be
\label{l_eft}
  {\cal L} = c_0 m^{3/2} X^{5/2} 
  + c_1 m^{1/2} \frac{(\vec{\nabla} X)^2}{\sqrt{X}} 
  + \frac{c_2}{\sqrt m} 
     \left[ \left(\nabla^2\varphi\right)^2 
           - 9m \nabla^2 V\right] \sqrt X\,,
\ee
where we have defined 
\be
  X = \mu - V - \dot\varphi 
    - \frac{(\vec{\nabla}\varphi)^2}{2m}\, .
\ee
Here, $\mu$ is the chemical potential and $V(\vec{x},t)$ is an
external potential. The functional form of the effective lagrangian 
is fixed by the symmetries of the problem, Galilean invariance, $U(1)$ 
symmetry, and conformal symmetry. The NLO effective lagrangian is 
characterized by three dimensionless parameters, $c_0,c_1,c_2$. These 
parameters can be related to physical properties of the system. The 
first parameter, $c_0$, can be related to the equation of state. We have 
\be
\label{c0}
  c_0 = \frac{2^{5/2}}{15\pi^2\xi^{3/2}}\,,
\ee
where $\xi$ determines the chemical potential in units of the 
Fermi energy, $\mu=\xi E_F$. The two NLO parameters $c_1,c_2$ are 
related to the momentum dependence of correlation functions. The 
phonon dispersion relation, for example, is given by
\be
\label{ph_disp}
  q_0 =  v_s q\left[ 1
   - \pi^2\sqrt{2\xi}\left(c_1+\frac{3}{2} c_2\right)
                \frac{q^2}{k_F^2} + O(q^4\log(q^2))\right]
\ee
where $v_s=\sqrt{\xi/3} v_F$ and $v_F=k_F/m$. The static susceptibility 
\be
 \chi(q) = -i\int dt\,d^3x\; e^{-i\vec{q}\cdot\vec{x}}\,
   \langle \psi^\dagger\psi(0) \psi^\dagger\psi(x)\rangle 
\ee
involves a different linear combination of $c_1$ and $c_2$, 
\be
\label{chi_q}
  \chi(q) = - \frac{mk_F}{\pi^2\xi} \left[
    1 + 2\pi^2\sqrt{2\xi}\left(c_1 - \frac{9}{2} c_2\right) 
    \frac{q^2}{k_F^2}+O(q^4\log(q^2))\right] .
\ee
Higher derivative terms in the effective lagrangian can also 
be used to compute the energy of inhomogeneous matter. At NLO
in an expansion in derivatives of the density we find the 
following energy density functional \cite{Rupak:2008xq}
\bea
{\cal E}(x) &=&  n(x)V(x) 
  + \frac{3\cdot 2^{2/3}}{5^{5/3}mc_0^{2/3}}n(x)^{5/3}
 - \frac{4}{45}\frac{2c_1+9c_2}{mc_0} 
     \frac{\left(\nabla n(x)\right)^2}{n(x)} \\ 
 & & \mbox{} - \frac{12}{5}\frac{c_2}{mc_0} \nabla^2 n(x) 
  \, . \nonumber 
\eea
The first two terms correspond to the local density approximation
(LDA) and the terms proportional to $c_1$ and $c_2$ are the leading 
correction to the LDA involving derivatives of the density. 

\begin{figure}[t]
\begin{center}
\includegraphics[width=7cm]{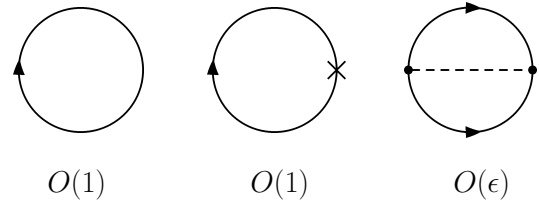}
\end{center}
\caption{\label{fig_veff}
Leading order contributions to the effective potential in the 
epsilon expansion. Solid lines denote fermions propagators, dashed
lines denote boson propagators, and the cross is an insertion of 
the chemical potential.  }
\end{figure}

\subsection{Epsilon Expansion}
\label{sec_eps}

 At unitarity the determination of $c_1$ and $c_2$ is a non-perturbative
problem, and we will perform the calculation using an expansion around
$d=4-\epsilon$ spatial dimensions \cite{Nussinov:2004,Nishida:2006br}.
Our starting point is the lagrangian 
\be
{\cal L}= \Psi^\dagger\left[
     i\partial_0+\sigma_3\frac{\vec\nabla^2}{2m}\right]\Psi
  + \mu\Psi^\dagger\sigma_3\Psi
  +\left(\Psi^\dagger\sigma_+\Psi\phi + h.c. \right)
  -\frac{1}{C_0}\phi^\dagger\phi\ ,
\ee
where $\Psi=(\psi_\uparrow,\psi_\downarrow^\dagger)^T$ is a two-component 
Nambu-Gorkov field, $\sigma_i$ are Pauli matrices acting in the Nambu-Gorkov 
space, $\sigma_\pm=(\sigma_1\pm i\sigma_2)/2$, $\phi$ is a complex boson
field, and $C_0$ is a coupling constant. In dimensional regularization 
the fermion-fermion scattering length becomes infinite for $1/C_0\to 0$. 

 The epsilon expansion is based on the observation that the fermion-fermion 
scattering amplitude near $d=4$ dimensions is saturated by the propagator 
of a boson with mass $2m$. The coupling of the boson to pairs of fermions
is given by
\be
   g  =\frac{\sqrt{8\pi^2\epsilon}}{m}
       \left(\frac{m\phi_0}{2\pi}\right)^{\epsilon/4} \, .
\ee
In the superfluid phase $\phi$ acquires an expectation value $\phi_0=
\langle\phi\rangle$. We write the boson field as $\phi = \phi_0 + g\varphi$. 
The lagrangian is split into a free part 
\be
{\cal L}_0 = \Psi^\dagger\left[i\partial_0+\sigma_3\frac{\vec\nabla^2}{2m}
     + \phi_0(\sigma_{+} +\sigma_{-})\right]\Psi
     + \varphi^\dagger\left(i\partial_0
        + \frac{\vec\nabla^2}{4m}\right)\varphi\, ,
\ee
and an interacting part ${\cal L}_I+{\cal L}_{ct}$, where  
\bea
{\cal L}_I &=& g\left(\Psi^\dagger\sigma_+\Psi\varphi + h.c\right)
     + \mu\Psi^\dagger\sigma_3\Psi  +2\mu \varphi^\dagger\varphi \, , \\
{\cal L}_{ct} &=& 
     - \varphi^\dagger\left(i\partial_0
        + \frac{\vec\nabla^2}{4m}\right)\varphi
     -2\mu \varphi^\dagger\varphi\, . 
\eea
Note that the leading self energy corrections to the boson propagator
generated by the interaction term ${\cal L}_I$ cancel against the 
counterterms in ${\cal L}_{ct}$. The chemical potential term for 
the fermions is included in ${\cal L}_I$ rather than in ${\cal L}_0$. 
This is motivated by the fact that near $d=4$ the system reduces to 
a non-interacting Bose gas and $\mu\to 0$. We will count $\mu$ as a 
quantity of $O(\epsilon)$. The Feynman rules are quite simple. The 
fermion and boson propagators are
\bea
\label{eps_prop}
G(p_0,p) &=& \frac{i}{p_0^2-E_{p}^2}
\left[\begin{array}{cc}
    p_0+\epsilon_{p} &  -\phi_0\\
    -\phi_0        & p_0-\epsilon_{p}
\end{array}\right]  \ ,\\
D(p_0, p) &=& \frac{i}{p_0-\epsilon_{p}/2}\ , 
\eea 
where $E_p^2=\epsilon_p^2+\phi_0^2$ and $\epsilon_p=p^2/(2m)$. The 
fermion-boson vertices are $ig\sigma^\pm$. Insertions of the 
chemical potential are $i\mu\sigma_3$. Both $g^2$ and $\mu$ are 
corrections of order $\epsilon$. 

\begin{figure}[t]
\begin{center}
\includegraphics[width=10cm]{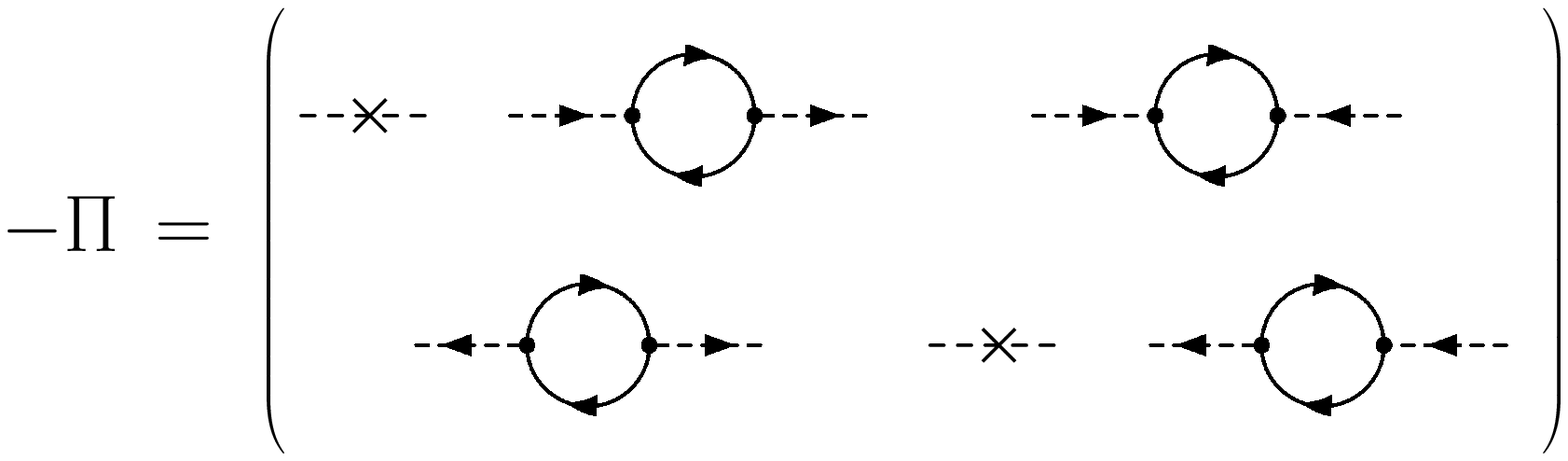}
\end{center}
\caption{\label{fig_sig_ph}
Scalar self energy at LO in the epsilon expansion. }
\end{figure}

\begin{figure}[t]
\begin{center}
\includegraphics[width=3cm]{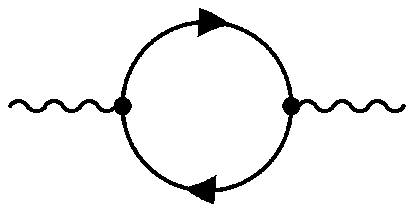}
\;\raisebox{0.038\vsize}{+}\;
\includegraphics[width=6cm]{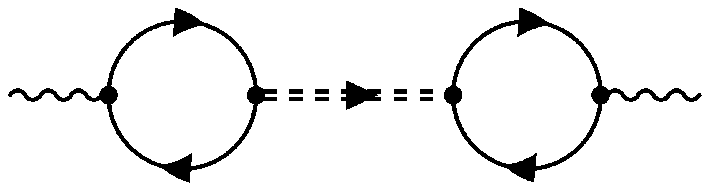}

\raisebox{0.038\vsize}{+}\;
\includegraphics[width=6cm]{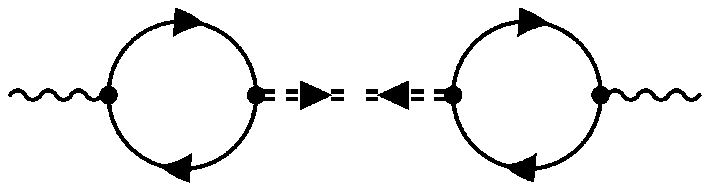}
\end{center}
\caption{\label{fig_sus}
Leading order contributions to the static susceptibility. The wiggly
line denotes an external current. The double line is the scalar
propagator defined in equ.~(\ref{ph_ds}). }
\end{figure}

 In order to determine $c_0,c_1,c_2$ we have to compute three
physical observables. We have studied $\xi=\mu/E_F$, and the 
curvature terms in the phonon dispersion relation and the static
susceptibility. The universal parameter $\xi$ was originally 
calculated by Nishida and Son. They computed the effective potential
to NLO in the epsilon expansion, see Fig.~\ref{fig_veff}. The
derivative of the effective potential with respect to $\mu$ determines
the density $n$, and the relation between $n$ and $\mu$ fixes $\xi$.
The result is
\be
\label{xi_eps}
\xi = \frac{\epsilon^{3/2}}{2}  \left[ 1  
      + \frac{1}{8}\epsilon \log(\epsilon)
      - \frac{1}{4}\left(12C-5+5\log(2)\right)\epsilon 
      + O(\epsilon^2) \right] \, , 
\ee
with $C=0.144$. The phonon dispersion relation can be extracted from the 
scalar propagator. We introduce a two-component scalar field $\Phi=(\varphi,
\varphi^*)$. The scalar propagator satisfies a Dyson-Schwinger 
equation \cite{Nishida:2006rp}
\be
\label{ph_ds}
\begin{array}{c}
\includegraphics[width=7cm]{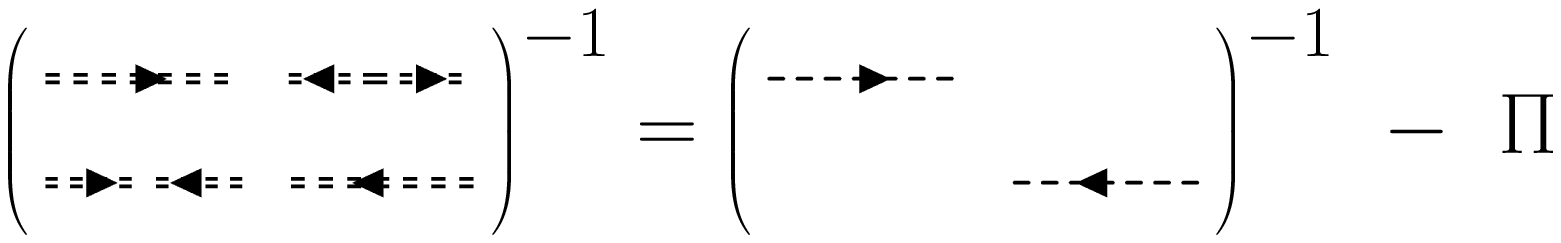}
\end{array}
\ee
At LO in the epsilon expansion the self energy is determined by
the diagrams shown in Fig.~\ref{fig_sig_ph}. NLO contributions
were calculated in \cite{Rupak:2008xq}. The phonon dispersion 
relation is 
\be 
\label{ph_disp_nlo}
 p_0 = \sqrt{\mu\epsilon_p} 
       \left(1+\frac{\epsilon}{8}\right) 
       \left\{ 1 +\frac{\epsilon_p}{8\mu} 
      \left(1-\frac{\epsilon}{4} \right)+ \ldots \right\}
\ee
We note that the dispersion relation curves up (unlike $^4$He, 
but similar to weakly interacting Bose gases). This implies that
there is $\varphi\to\varphi+\varphi$ decay. Finally, we can determine 
the static susceptibility. Computing the diagrams in Fig.~\ref{fig_sus} 
we get \cite{Rupak:2008xq,Kryjevski:2008si}
\bea 
\label{chi_nlo}
\chi(q) &=& -\frac{2Z}{\epsilon\mu} 
 \left\{ 1 - \frac{1}{8}\left(\frac{q^2}{m\mu}\right) 
      \left(1-\frac{\epsilon}{4} \right)+ O(q^4) \right\}
    \left(\frac{m\phi_0}{2\pi}\right)^{d/2}\, ,  \\
 & & \mbox{} 
 Z\,=\,  1 - \frac{1}{2}\left(\gamma-\log(2)\right)\epsilon \, . 
   \nonumber 
\eea
The coefficient $c_0$ follows from the result for $\xi$ ($\xi=0.475$
at NLO in the $\epsilon$-expansion) using equ.~(\ref{c0}). Matching 
equ.~(\ref{ph_disp_nlo},\ref{chi_nlo}) against equ.~(\ref{ph_disp},\ref{chi_q})
gives $c_2=0$ and $c_1/c_0=3/8-\epsilon/4$. The corresponding energy 
density functional was studied in \cite{Rupak:2008xq}. Compared to a 
free Fermi gas the local density term is reduced by a factor $\sim 2$ 
(the interaction is attractive), while the gradient correction proportional 
to $(\nabla n)^2/n$ is enhanced by a factor $\sim 2$. 

\section{Transport Properties}
\label{sec_tra}

 In the following we will discuss transport properties of the 
Fermi gas at unitarity. The interest in non-equilibrium properties
arises from the observation that transport coefficients are much
more sensitive to the strength of the interaction than thermodynamic
quantities. A renewed interest in transport properties was also 
sparked the AdS/CFT correspondence and the experimental limits on
the shear viscosity of the quark gluon plasma obtained at RHIC. In 
the following we shall focus on the shear viscosity of the Fermi 
gas at unitarity. Close to equilibrium the (coarse grained) energy momentum 
tensor can be written as
\bea
\label{del_T}
T_{ij}        &=& (P+\epsilon)v_i v_j -P\delta_{ij} +\delta T_{ij}\ ,\\
\delta T_{ij} &=&-\eta(\nabla_i v_j+\nabla_i
v_j-\frac{2}{3}\delta_{i j}\vec{\nabla}\cdot\vec{v})+\cdots ,\nonumber
\eea
where $\epsilon$ and $P$ are the energy density and pressure, and 
$v_i$ is the local flow velocity. The first term is the ideal gas
contribution, and $\delta T_{ij}$ is the leading order (in gradients
of $v_i$) dissipative correction. The traceless part of $\delta T_{ij}$
is proportional to the shear viscosity $\eta$.
  
\subsection{Kinetic Theory}
\label{sec_kin}

\begin{figure}[t]
\begin{center}
\raisebox{0.085\vsize}{a)}\;
\includegraphics[width=6cm]{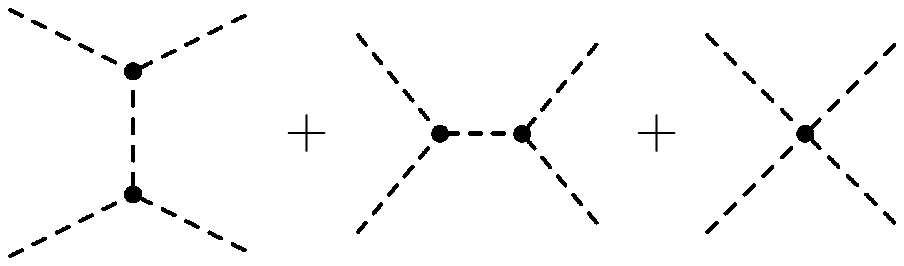}
\hspace*{1cm}
\raisebox{0.085\vsize}{b)}\;
\includegraphics[width=1.6cm]{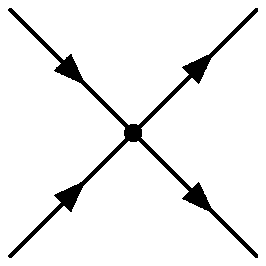}
\end{center}
\caption{\label{fig_kin}
Leading order processes that contribute to the shear viscosity at 
low temperature (Fig.~a) and high temperature (Fig.~b). Dashed lines
are phonon propagators and solid lines are fermion propagators. }
\end{figure}

 We first consider the case that the fluid is composed of weakly 
interacting quasi-particles. In the unitarity limited Fermi gas 
this is the case at $T\ll T_c$ (phonons) and $T\gg T_c$ (atoms). 
In these limits we can compute the shear viscosity using kinetic
theory. In the following we will concentrate on the low temperature 
case discussed in \cite{Rupak:2007vp}. In kinetic theory the stress-energy 
tensor is given by 
\be
\label{Tij_kin}
T_{ij} = v_s^2\int\frac{d^3p}{(2\pi)^3}
   \frac{p_i p_j}{E_p} f_p\, ,
\ee
where $f_p$ is the distribution function of the phonons, $v_s$ is the 
speed of sound, $p_i$ is the momentum and $E_p$ the quasi-particle energy. 
Close to equilibrium $f_p=f_p^{(0)}+\delta f_p$, where $f_p^{(0)}$ is 
the Bose-Einstein distribution and $\delta f_p$ is a small departure from 
equilibrium. We write $\delta f_p=-\chi(p) f_p^{(0)}(1+f_p^{(0)})/T$. In
the case of shear viscosity we can further decompose
\be
\label{ansatz}
\chi(p) = g(p) (p_i p_j-\frac{1}{3}\delta_{i j}p^2)
  (\nabla_i v_j+\nabla_j v_i
     -\frac{2}{3} \delta_{i j}\vec{\nabla}\cdot\vec{v}) \, . 
\ee
Inserting equ.~(\ref{ansatz}) into equ.~(\ref{Tij_kin}) we get
\be
\label{viscosity}
\eta =\frac{4v^2}{15 T }
\int \frac{d^3 p}{(2\pi)^3}
\frac{p^4}{2 E_p}f_p^{(0)}(1+f_p^{(0)}) g(p) \, . 
\ee
The non-equilibrium distribution $g(p)$ is determined by the Boltzmann
equation 
\be
\label{boltzmann}
\frac{d f_p}{dt}=\frac{\partial f_p}{\partial t}+ \vec{v}\cdot
\vec{\nabla} f_p = C[f_p],
\ee
relating the rate of change of the distribution function $f_p$ to the
collision operator $C[f_p]$. The  $2\leftrightarrow2$ collision integral 
is given by 
\bea
C_{2\leftrightarrow2}[f_p]&=&\frac{1}{2 E_p}\int \frac{d^3 k}{(2\pi)^3 2
  E_k}
\frac{d^3 k'}{(2\pi)^3 2 E_{k'}}
\frac{d^3 p'}{(2\pi)^3 2 E_{p'}}\\
& &\times(2\pi)^4\delta^{(4)}(p+k-p'-k')|{\cal M}|^2
D_{2\leftrightarrow2}, \nonumber 
\eea
where $D_{2\leftrightarrow 2}$ contains the distribution functions
and $|{\cal M}|$ is the $2\leftrightarrow 2$ scattering amplitude 
shown in Fig.~\ref{fig_kin}. The three and four-phonon vertices are
fixed by the effective lagrangian (\ref{l_eft}). Linearizing 
$D_{2\leftrightarrow 2}$ in $\delta f_p$ one finds
\be
 D_{2\leftrightarrow2} =  \frac{1}{T}
  f_{k'}^{(0)} f_{p'}^{(0)} (1+f_k^{(0)})(1+f_p^{(0)}) \, 
 \left( \chi(p)+\chi(k)- \chi(p')-\chi(k')\right)\, .
\ee
There are a variety of methods for solving the linearized Boltzmann 
equation. A standard technique is based on expanding $g(p)$ in a
complete set of functions. A nice feature of this method is that the
truncated expansion gives a variational estimate  
\be
\eta \geq \frac{4v^4}{25 T^2} 
  \frac{(b_0A_{00})^2}{\sum_{s, t} b_s b_t M_{s t}} \,
\ee
where $b_s$ is a set of expansion coefficients, $A_{00}$ is a normalization
integral, and $M_{st}$ are matrix elements of the linearized collision 
operator. For the best trial function we find \cite{Rupak:2007vp}
\be
\eta/s =
7.7\times 10^{-6}\xi^5\frac{T_F^8}{T^8}\, , 
\ee
where $\xi$ is the universal parameter introduced in Sect.~\ref{sec_let}
and we have normalized the result to the entropy density $s$ of a 
weakly interacting phonon gas. A similar estimate can be obtained 
in the high temperature limit. In this case the relevant degrees of 
freedom are atoms, and the dominant scattering process is shown in 
Fig.~\ref{fig_kin}b. The result is \cite{Bruun:2005,Bruun:2007}
\be 
\eta/s = \frac{45\pi^{3/2}}{64\sqrt{2}} 
\left(\frac{T}{T_F}\right)^{3/2}
\left[\log\left(\frac{3\sqrt{\pi}}{4} 
\frac{T^{3/2}}{T_F^{3/2}}\right)+\frac{5}{2}\right]^{-1} \, .
\ee
The high and low temperature limits of $\eta/s$ are shown in 
Fig.~\ref{fig_eta_s}, together with the proposed lower bound
$\eta/s=1/(4\pi)$ \cite{Kovtun:2004de} and experimental data
which we will discuss in the next section. 

\subsection{Hydrodynamics}
\label{sec_hydro}

\begin{figure}[t]
\begin{center}
\includegraphics[width=8cm]{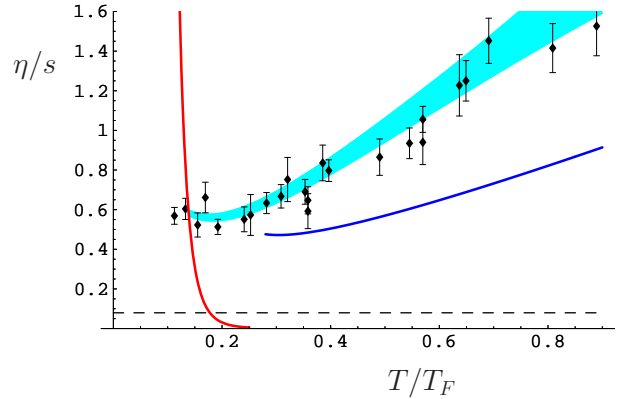}
\end{center}
\caption{\label{fig_eta_s}
Viscosity to entropy density ratio of a cold atomic gas in the
unitarity limit. This plot is based on the damping data
published in \cite{Kinast:2005} and the thermodynamic data in
\cite{Kinast:2005b,Luo:2006}. The dashed line shows the
conjectured viscosity bound $\eta/s=1/(4\pi)$, and the solid
lines show the high and low temperature limits.  }
\end{figure}

 Hydrodynamics describes the evolution of long-wavelength, slow-frequency
modes. The hydrodynamic description remains valid even if there is no
underlying kinetic theory. The hydrodynamic equations follow from
conservation of mass (particle number), energy and momentum. In 
a non-relativistic system the equations of continuity and of momentum 
conservation are given by 
\bea
\frac{\partial n}{\partial t} + \vec{\nabla}\cdot\left(n\vec{v}\right) 
 &=& 0 , \\
mn \frac{\partial \vec{v}}{\partial t} 
 + mn \left(\vec{v}\cdot\vec{\nabla} \right)\vec{v} &=& 
 -\vec{\nabla}P-n\vec{\nabla}V,
\eea 
where $n$ is the number density, $m$ is the mass of the atoms, $\vec{v}$ 
is the fluid velocity, $P$ is the pressure and $V$ is the external
potential. In an ideal fluid the equation of energy conservation can 
be rewritten as conservation of entropy, 
\be  
\frac{\partial ns}{\partial t} + \vec{\nabla}\cdot\left(ns\vec{v}\right) 
 = 0\, . 
\ee
A non-zero shear viscosity leads to dissipation, converting kinetic
energy to heat and increasing the entropy. The shear viscosity of 
the dilute Fermi gas in the unitarity limit can be measured by studying 
the damping of collective modes in trapped systems \cite{Kinast:2004}. 
The frequency of these modes agrees well the prediction of ideal 
hydrodynamics. The dissipated energy is given by
\be
\label{E_dot}
\dot{E} = -\frac{1}{2} \int d^3x\, \eta(x)\, 
  \left(\partial_iv_j+\partial_jv_i-\frac{2}{3}\delta_{ij}
      \partial_k v_k \right)^2  \, . 
\ee
The damping rate is given by the ratio of the energy dissipated to 
the total energy of the collective mode. The kinetic energy is 
\be 
\label{E_kin}
 E_{kin} = \frac{m}{2}\, \int d^3x\, n(x) \vec{v}^{\,2} \, . 
\ee
If the damping rate is small both $\dot{E}$ and $E_{kin}$ can be 
computed using the solution of ideal hydrodynamics. We recently 
performed an analysis \cite{Schafer:2007pr} which is based on 
measurements of the damping rate of the lowest radial breathing
mode performed by the Duke group \cite{Kinast:2005}. We showed 
that can relate the dimensionless ratio $\Gamma/\omega$, where 
$\Gamma$ is the damping rate and $\omega$ is the trap frequency, 
to the shear viscosity to entropy density 
ratio
\be 
\label{eta_s}
\frac{\eta}{s}  =\frac{3}{4} \xi^{1/2} (3N)^{1/3} 
 \left(\frac{\bar\omega\Gamma}{\omega_\perp^2}\right)
 \left(\frac{E}{E_{T=0}}\right)
 \left(\frac{N}{S}\right).
\ee
Here $N$ is the total number of particles in the trap ($2\cdot 10^5$
in \cite{Kinast:2005}), $\xi$ is the universal parameter defined 
in Sec.~\ref{sec_let}, $E/E_{T=0}$ is the ratio of the total energy
to the energy at $T=0$ (which can be extracted using a Virial theorem
from the measured cloud size), and $S/N$ is the entropy per particle
(which is measured using adiabatic sweeps to the BCS limit \cite{Luo:2006}). 
The results are compared to theoretical prediction in the high and low 
temperature limit in Fig.~\ref{fig_eta_s}. The data show a minimum near 
$T/T_F\simeq 0.2$. At the minimum $\eta/s\sim 1/2$. This should probably 
be considered as an upper bound, since dissipative mechanism other than 
shear viscosity may be present. In the high $T$ limit there is fairly 
good agreement with kinetic theory. The temperature dependence implied 
by the low $T$ prediction is not seen in the data. This is maybe not 
very surprising, since the mean free path in the low $T$ regime quickly 
exceeds the system size. 

\section{Outlook}
\label{sec_sum}

 There are many promising directions for further study. Clearly,
it is desirable to obtain additional experimental constraints
on the shear viscosity, and to improve the theoretical analysis
of the existing data sets. It would also be interesting to 
confirm that the bulk viscosity vanishes in the normal phase, 
and to measure the thermal conductivity. We would also like
to improve the theoretical tools for computing transport 
properties in the interesting regime near $T_c$. There are 
some recent ideas for applying holography and the $AdS/CFT$
correspondence to Galilean invariant conformal field theories
\cite{Son:2008ye,Balasubramanian:2008dm}, but there are also
many purely field theoretic methods ($\epsilon$ expansions, 
large $N$ methods) that have yet to be pursued.
 
Acknowledgments: Much of this work was carried out in collaboration
with G.~Rupak. The work is supported in part by the
US Department of Energy grant DE-FG02-03ER41260.


\end{document}